\documentclass[hyper]{JHEP3}

\title{Infrared finite coupling in Sudakov resummation}
\author{G.\ Grunberg\\
        Centre de Physique Th\'eorique de l' Ecole  
Polytechnique (CNRS UMR C7644),\\
        91128 Palaiseau Cedex, France\\
        E-mail: \email{georges.grunberg@pascal.cpht.polytechnique.fr}}
\abstract{New arguments are presented to emphasize the interest of the infrared finite coupling approach to power
corrections in the context of Sudakov resummation. The more regular infrared behavior of some peculiar
combinations of Sudakov anomalous dimensions, free of Landau singularities at large $N_f$, is pointed out. A
general conflict between the infrared finite coupling and infrared renormalon approaches to power corrections is
explained, and a possible resolution is proposed, which makes use of the arbitrariness of the choice of
exponentiated constant terms. A simple ansatz for a 'universal' non-perturbative Sudakov effective coupling at
large $N_f$ follows naturally from these considerations. In this last version, a new result is presented: the
striking emergence of an infrared finite {\em perturbative} effective coupling in the Drell-Yan process at large
$N_f$ (at odds with the infrared renormalon argument) within the framework of Sudakov resummation for eikonal cross
sections of Laenen, Sterman and Vogelsang. Some suggestions for phenomenology at finite
$N_f$, alternative to the shape function approach, are given.}
\preprint{CPTh/PC 006.0106}
\begin{document}

\section{Introduction}
The notion of an infrared (IR) finite coupling, and the related concept of universality,  to parametrize power
corrections in QCD has attracted much attention for a long time \cite{DMW}, \cite{DW}. In the present note,
which is an extended version of a talk given at the  FRIF workshop on non-perturbative effects in jets (Paris,
January 10-14 2006), I display further evidence in favor of this assumption in the more specific framework of
Sudakov resummation. A more detailed account of the topics covered here is postponed to a future publication. 
The paper is organized as follows. Section 2 gives a phenomenological incentive for an IR finite effective coupling
to parametrize the tail of the Sudakov peak for the thrust distribution. Section 3 gives a theoretical incentive,
where some remarkable IR properties of specific combinations of Sudakov anomalous dimensions are pointed out.
Section 4 discusses a general conflict between the renormalon and the IR finite coupling approaches to power
corrections, and indicates a possible resolution based on the arbitrariness of the choice of constant terms
exponentiated in Sudakov resummation along with the large logarithms. This issue is further discussed  in sections
5-7, and a closed form solution is given at large
$N_f$ in the case of deep inelastic scattering (DIS) in section 7.1. An application to the Drell-Yan process is
given in section 7.2, where it is shown that the exponentiation of ${\cal O}(N^0)$ terms advocated in
\cite{Vogelsang} naturally leads to the emergence of the simplest ansatz for an IR finite {\em perturbative}
coupling (albeit being at odds with the IR renormalon argument). Alternatively, an  attempt to reconcile the IR
renormalon and IR coupling approaches leads to another  ansatz for a `universal', but non-perturbative, Sudakov
effective coupling at  large
$N_f$  proposed in section 8. Section 9 deals further with the issue of resummation of constant terms, and
sketches a procedure for phenomenology at finite $N_f$.  Section 10 contains the conclusions.

\section{Incentive for IR finite coupling approach}

Let us take as an example the case of thrust, and  define

\begin{equation}{1\over \sigma_{tot}}\int_{0}^{\tau_{max}}d\tau\ \exp(-\nu \tau)\ {d\sigma
\over d\tau}\equiv\exp\left[E(Q^2,\nu)\right] \label{eq:exp-thrust},\end{equation}
where $\tau\equiv 1-T$. Consider now the case where $\nu$ is large, which corresponds to $\tau$ small, and
assume  the standard \cite{CTTW} exponentiation formula

\begin{equation}E(Q^2,\nu)\equiv S(Q^2,\nu)+H(Q^2)+{\cal O}(1/\nu)\label{eq:E},\end{equation}
where $ H(Q^2)$ is a power series in $a_s(Q^2)\equiv \alpha_s(Q^2)/4\pi$ with coefficients independent of $\nu$,
and the `Sudakov exponent'
$S(Q^2,\nu)$ is given by

\begin{equation}S(Q^2,\nu)=\int_{0}^1 {dx \over x}\left[\exp(-\nu x)-1\right]\left[4 C_F\int_{x^2 Q^2}^{x Q^2}
{dk^2
\over k^2}\ A_{{\cal S}}(k^2)-3 C_FB_{{\cal S}}(x Q^2)\right]
\label{eq:Sud-thrust},\end{equation}
where $A_{{\cal S}}(k^2)$ and $B_{{\cal S}}(k^2)$, the `Sudakov effective couplings', should be considered as
two physical `effective charges' \cite{GG} which can be expanded as power series in $a_s(k^2)$

\begin{equation}A_{{\cal S}}(k^2)=a_s(k^2)+{\cal A}_1 a_s^2(k^2)+{\cal A}_2 a_s^3(k^2)+...
\label{eq:AS-PT},\end{equation}
and similarly for $B_{{\cal S}}(k^2)$.
It is convenient to 
interchange the $x$ and $k^2$ integrations, and write eq.(\ref{eq:Sud-thrust}) as a sum of
two `renormalon integrals' \cite{Gru-power}

\begin{equation}S(Q^2,\nu)=4 C_F\int_{0}^{Q^2}{dk^2\over k^2} F_A(k^2/Q^2,\nu) A_{{\cal S}}(k^2) -3 C_F
\int_{0}^{Q^2}{dk^2\over k^2}F_B(k^2/Q^2,\nu) B_{{\cal S}}(k^2)
\label{eq:ren-int},\end{equation}
where $F_A(k^2/Q^2,\nu)$ and
$F_B(k^2/Q^2,\nu)$, the `Sudakov distribution functions', are given by

\begin{equation}F_A(k^2/Q^2,\nu)=\int_{k^2/Q^2}^{k/Q} {dx \over
x}\left[\exp(-\nu x)-1\right]\label{eq:F-A},\end{equation} and

\begin{equation}F_B(k^2/Q^2,\nu)=\exp(-\nu k^2/Q^2)-1\label{eq:F-B}.\end{equation}
I note that

\begin{equation}F_A(k^2/Q^2,\nu)\rightarrow \ln(k/Q) \label{eq:F-A-Nlarge}\end{equation}
and

\begin{equation}F_B(k^2/Q^2,\nu)\rightarrow -1 \label{eq:F-B-Nlarge}\end{equation}
for $\nu\rightarrow \infty$.
Assume now that the effective couplings 
$A_{{\cal S}}(k^2)$ and $B_{{\cal S}}(k^2)$ are IR finite, and setting
$A_{{\cal S}}(k^2)\equiv A_{IR}+\Delta A_{{\cal S}}(k^2)$ and $B_{{\cal S}}(k^2)\equiv B_{IR}+\Delta B_{{\cal
S}}(k^2)$, assume further that
$\Delta A_{{\cal S}}(k^2)$ and $\Delta B_{{\cal S}}(k^2)$ are ${\cal O}(k^2)$ for $k^2\rightarrow 0$. Then,
taking the
$\nu\rightarrow\infty$ limit  under the integrals, one gets 

\begin{eqnarray}S(Q^2,\nu)&\sim&  -4 C_F K_1(\nu) A_{IR}-3 C_F K_0(\nu) B_{IR}\nonumber\\
&+&4 C_F\int_{0}^{Q^2}{dk^2\over k^2}
\ln(k/Q)
\ \Delta A_{{\cal S}}(k^2)+3 C_F\int_{0}^{Q^2}{dk^2\over k^2}\Delta B_{{\cal
S}}(k^2)\label{eq:Sud-Nlarge},\end{eqnarray} 
where the integrals over $\Delta A_{{\cal S}}(k^2)$ and $\Delta B_{{\cal S}}(k^2)$ are $\nu$-independent (and
IR convergent), whereas 
\begin{equation}K_1(\nu)=\int_{0}^1 {dx \over x}\left[\exp(-\nu x)-1\right]\ln x\sim 1/2\
\ln^2\nu\label{eq:K11}\end{equation} 
and
\begin{equation}K_0(\nu)=\int_{0}^1 {dx \over x}\left[\exp(-\nu x)-1\right]\sim -\ln\nu\label{eq:K00}\end{equation}  
for
$\nu\rightarrow\infty$. Thus we obtain at large $\nu$

\begin{equation}E(Q^2,\nu)\sim -4 C_F K_1(\nu) A_{IR}-3 C_F K_0(\nu) B_{IR}+{\cal
C}(Q^2)\label{eq:E-Nlarge},\end{equation} where

\begin{equation}{\cal C}(Q^2)\equiv 4 C_F\int_{0}^{Q^2}{dk^2\over k^2} \ln(k/Q)
\ \Delta A_{{\cal S}}(k^2)+3 C_F\int_{0}^{Q^2}{dk^2\over k^2}\Delta
B_{{\cal S}}(k^2)+H(Q^2)\label{eq:constant}\end{equation} is $\nu$-independent.
Hence for $\nu\rightarrow\infty$

\begin{equation}{1\over \sigma_{tot}}\int_{0}^{\tau_{max}}d\tau\ \exp(-\nu \tau)\ {d\sigma
\over d\tau}\sim {\cal N}(Q^2)\ \exp\left[-2  C_F A_{IR}\ln^2\nu+{\cal O}(\ln\nu)\right]
\label{eq:tail},\end{equation}
where ${\cal N}(Q^2)$ is a non-perturbative $Q$-dependent normalization constant (its $Q$-dependence can
eventually be estimated at large $Q$).  Upon taking the inverse Laplace transform the present ansatz  can thus
potentially (if
$A_{IR}>0$) reproduce the expected tail of the thrust distribution for $\tau\rightarrow 0$, which is
parametrized essentially by the IR value $A_{IR}$ of the effective coupling $A_{{\cal S}}(k^2)$ ($B_{IR}$
gives a subdominant contribution) which could be fitted from the data. I note that no assumption that $Q$ is
large has been made, so this prediction goes {\em beyond} short distance physics. To fit the whole thrust
distribution at all values of
$\tau$, one should make a more complete ansatz for the IR behavior of $A_{{\cal S}}(k^2)$ and $B_{{\cal
S}}(k^2)$. This step requires a priori no more free parameters than the `shape function'
approach \cite{Korchemsky}  and provides an alternative to it.

\section{IR behavior of the Sudakov effective  coupling in the $N_f\rightarrow \infty$ limit}

Let us take as a case study the example of the scaling violation for the non-singlet structure function in
deep inelastic scattering (DIS). From the standard exponentiation formulas in Mellin $N$- space (I shall adopt in
general the notations of \cite{Vogt}), one gets immediately the relation at large $N$ for the `physical
anomalous dimension' \cite{GG,Catani} $d\ln F_2(Q^2,N)/d\ln Q^2$

\begin{equation}{d\ln F_2(Q^2,N)\over d\ln Q^2}=4 C_F\ H(Q^2)+4 C_F\int_{0}^1 dz{z^{N-1}-1 \over
1-z} A_{{\cal S}}[(1-z)Q^2]+{\cal O}(1/N)  
\label{eq:scale-viol},\end{equation}
where $H(Q^2)$ is again given as a power series in
$a_s(Q^2)$ with $N$-independent coefficients
and

\begin{equation}4 C_F A_{{\cal S}}(k^2)= A(a_s(k^2))+{dB(a_s(k^2))\over d\ln
k^2}\label{eq:S-coupling}.\end{equation}
$A$ (the universal  `cusp' anomalous dimension) and $B$ are  the standard Sudakov anomalous dimensions
relevant to DIS, given as  power series in
$a_s$: $A(a_s)=A_0 a_s+A_1 a_s^2+..$ (with $A_0=4 C_F$) and
$B(a_s)=B_0 a_s+B_1 a_s^2+..$. I assumed the conjecture of \cite{Gardi-Roberts}, relative to the vanishing to
all orders of the third standard anomalous dimension $D$ (the `soft function'), is correct.
$A_{{\cal S}}=a_s+{\cal O}(a_s^2)$ shall be refered to as the `Sudakov effective coupling', and we
 shall see (in the large $N_f$ limit) that it has drastically different infrared properties then the individual
Sudakov anomalous dimensions it is composed of. 

\noindent\underline{Large $N_f$ analysis}: the Borel transform $B[A_{{\cal S}}](u)$  of the Sudakov
effective coupling has been computed at large $N_f$. It is defined by

\begin{equation} A_{{\cal S}}(k^2)=\int_0^\infty du\
\exp\left(-u\ln{k^2 \over\Lambda^2}\right)\
 B[A_{{\cal S}}](u) \label{eq:Borel-coupling1},\end{equation}
and is was found \cite{Gardi,Gardi-Roberts} (this result is checked below using a different method) that at
large $N_f$, with the `na\"{\i}ve non-abelization' recipie \cite{Beneke}

\begin{equation}B[A_{{\cal S}}](u)={1\over \beta_0}\exp(-d u){\sin\pi u\over \pi u}\ {1\over
2}\left({1\over 1-u}+{1\over 1-u/2}\right) 
\label{eq:Borel-coupling2},\end{equation}
where $\beta_0={11\over 3} C_2(G)-{2\over 3}N_f$ is the one-loop coefficient of the
beta function and $d$ is a scheme-dependent constant related to the renormalization of fermion loops: $d=-5/3$
in the $\overline{MS}$ scheme and $d=0$ in the so called `V-scheme' (or `single dressed gluon scheme'). In the
following I shall use the `V-scheme' for simplicity. One should note that $B[A_{{\cal S}}](u)$  is free of
renormalons singularities, and in fact the corresponding standard perturbative series of
$A_{{\cal S}}$ in powers of $a_s$ has a finite convergence radius.

One first observes that for $k=\Lambda$ (the Landau pole of the one-loop V-scheme coupling) the Borel integral
in eq.(\ref{eq:Borel-coupling1}) converges and
$A_{{\cal S}}(k^2=\Lambda^2)$ is well-behaved and finite. This behavior is in striking contrast with that of
the cusp anomalous dimension, which has also been computed \cite{Beneke-Braun}, \cite{Gracey}  at large $N_f$ 

\begin{equation}A(a_s)=4 C_F {\sin \pi\beta_0 a_s\over\pi\beta_0} {\Gamma(4+2\beta_0 a_s)\over
6\Gamma(2+\beta_0 a_s)^2}
\label{eq:cusp},\end{equation}
where  $\beta_0 a_s=1/\ln k^2/\Lambda^2$ at large $N_f$. It is easy to see, since $a_s\rightarrow\infty$ for 
$k\rightarrow\Lambda$, that the ratio of gamma functions in eq.(\ref{eq:cusp}) blows up in this limit, while
the sin factor generates wild oscillations, resulting in a completely unphysical behavior around $k=\Lambda$.
The much more tamed behavior of $A_{{\cal S}}(k^2)$ suggests that it is well-behaved in the infrared
region.  It is actually possible to get an analytic expression for
$A_{{\cal S}}(k^2)$, valid at all $k^2$. One finds

\begin{eqnarray}A_{{\cal S}}(k^2)&=&A_{{\cal S}}^{simple}(k^2)+{1\over
2\pi\beta_0}\exp(-t){Ei(t+i\pi)-Ei(t-i\pi)\over 2i}\nonumber\\
&-&{1\over
2\pi\beta_0}\exp(-2t){Ei(2t+2i\pi)-Ei(2t-2i\pi)\over 2i}
\label{eq:S-coupling1},\end{eqnarray}
with

\begin{equation}A_{{\cal S}}^{simple}(k^2)={1\over\beta_0}\left[{1\over
2}-{1\over\pi}\arctan(t/\pi)\right]\label{eq:A-simple},\end{equation}
where $t=\ln k^2/\Lambda^2$ and $Ei(x)$ is the exponential integral function.
Eq.(\ref{eq:S-coupling1}) can equivalenly be written in term of the incomplete gamma function
$\Gamma(0,x)=\int_x^{\infty}{dz\over z} \exp(-z)$ as

\begin{eqnarray}A_{{\cal S}}(k^2)&=&A_{{\cal S}}^{simple}(k^2)+{1\over
2\pi\beta_0}\exp(-t)\left[{\Gamma(0,-t+i\pi)-\Gamma(0,-t-i\pi)\over 2i}+\pi\right]\nonumber\\
&-&{1\over
2\pi\beta_0}\exp(-2t)\left[{\Gamma(0,-2t+2i\pi)-\Gamma(0,-2t-2i\pi)\over 2i}+\pi\right]
\label{eq:S-coupling2}.\end{eqnarray}
To investigate the infrared behavior, I now make use of the fact that
$\Gamma(0,x)\simeq {\exp(-x)\over x}(1-1/x+...)$ for $\vert x\vert\rightarrow\infty$, which implies

\begin{eqnarray}\exp (-t)\left[{\Gamma(0,-t+i\pi)-\Gamma(0,-t-i\pi)\over 2i}\right]&\simeq&{\pi\over
t^2}\nonumber\\
\exp (-2t)\left[{\Gamma(0,-2t+2i\pi)-\Gamma(0,-2t-2i\pi)\over 2i}\right]&\simeq&-{2\pi\over 4t^2}
\label{eq:asymp}\end{eqnarray}
for $\vert t\vert\rightarrow\infty$. In the ultraviolet (UV) region ($t\rightarrow +\infty$) one thus obtain, as
expected, a power series in $a_s(k^2)=1/\beta_0 t$

\begin{equation}A_{{\cal S}}(k^2)\simeq a_s(k^2)+{3\over 4}\beta_0\
a_s^2(k^2)+({5\over 4}-{\pi^2\over 3})\beta_0^2\ a_s^3(k^2)+...\label{eq:couplingUV}.\end{equation} 
I note that $A_{{\cal S}}^{simple}(k^2)$    behaves as
$a_s+{\cal O}(a_s^3)$ in this limit, while the last two terms on the
right hand side of eq.(\ref{eq:S-coupling2}) are ${\cal O}(a_s^2)$. $A_{{\cal S}}^{simple}(k^2)$ turns
out to coincide for real space-like
$k^2=\mu^2>0$ with the (integrated) time-like discontinuity at $k^2=-\mu^2<0$ of the one-loop
coupling. I stress however that here it should be considered as an analytic contribution to the
large
$N_f$ Sudakov effective coupling, an {\em Euclidean} coupling defined in the whole complex $k^2$ plane.
The analytic continuation to complex
$k^2$ is conveniently provided by  the identity:

\begin{equation}{1\over 2}-{1\over \pi}\arctan(t/\pi)={1\over \pi}{\ln(t+i \pi)-\ln(t-i \pi)\over
2i}\label{eq:couplingnaive}.\end{equation} 
On the other hand, in the IR region ($t\rightarrow -\infty$) $A_{{\cal S}}^{simple}(k^2)$  reaches a {\em
finite} limit of
$1/\beta_0$. Since the two incomplete gamma function contributions are ${\cal O}(1/t^2)$, and thus also vanish
for $t\rightarrow -\infty$, the full $A_{{\cal S}}(k^2)$ would have reached the {\em same}  IR fixed
point value of
$1/\beta_0$, were it not for the two IR divergent contributions $\pi\exp(-t)$ and $\pi\exp(-2t)$. Thus in fact
$A_{{\cal S}}(k^2)$ approaches an {\em infinite} (and negative) IR fixed point for $k^2\rightarrow 0$: 
 
\begin{equation}A_{{\cal S}}(k^2)\simeq -{1\over 2\beta_0}{\Lambda^4\over k^4}+{1\over
2\beta_0}{\Lambda^2\over k^2}
\label{eq:couplingIR}.\end{equation}
This is not by itself  an unphysical behavior, except
for  the  negative sign in the infrared, which cannot reproduce a vanishing Sudakov tail. Moreover $A_{{\cal
S}}(k^2)$ is a {\em causal} function and  has no unphysical Landau singularities on the whole first sheet of the
$k^2$ plane: the only possible branch points occur for
$t=\pm i\pi$, i.e. at $k^2/\Lambda^2=-1$ on the time-like axis. The  trouble  is in the too strongly
divergent IR behavior, which gives a  divergent contribution ${\Lambda^4\over Q^4}{\cal O}({1\over(1-z)^2}$)
for
$z\rightarrow 1$ to the integral on the right-hand side of eq.(\ref{eq:scale-viol}). This finding makes it
more plausible the speculation \cite{DMW} that there exists a non-perturbative modification $\delta
A_{{\cal S},NP}(k^2)$ of the coupling at very small momenta which might turn the too strong IR
fixed point of perturbative origin into a genuinely non-perturbative, but softer (eventually finite) IR fixed
point, yielding an IR convergent Sudakov integral.  A simple ansatz for $\delta A_{{\cal S},NP}(k^2)$,
localized in the IR region, and which thus does not induce large OPE-violating \cite{DMW}  ultraviolet ${\cal
O}(1/Q^2)$ power corrections, is given by\footnote{The straightforward modification of the coupling obtained
by just subtracting the IR divergent contribution eq.(\ref{eq:couplingIR}) yields, when viewed from the UV
side, a too large short distance ${\cal O}(1/Q^2)$ correction.}

\begin{equation}\delta A_{{\cal S},NP}(k^2)={1\over 2\beta_0}\left[-{\Lambda^2\over
k^2}\exp\left(-{k^2\over \Lambda^2}\right)+{\Lambda^4\over k^4}\exp\left(-{k^4\over \Lambda^4}\right)\right]
\label{eq:deltacoupling}.\end{equation}
Then

\begin{equation}A_{{\cal S},NP}(k^2)=A_{{\cal S}}(k^2)+\delta A_{{\cal S},NP}(k^2)
\label{eq:npcoupling},\end{equation}
which indeed yields for $k^2\rightarrow 0$  a finite and positive IR fixed point:
$A_{{\cal S},NP}(k^2)\simeq 1/\beta_0+{\cal O}(k^2)$.

\section{A clash between the infrared finite coupling  and the IR renormalons approaches to power
corrections}
There is a potential clash between the infrared finite coupling  and the IR renormalons approaches to power
corrections (very closely connected to the well-known  issue \cite{GS},  \cite{Beneke-Braun}  of $1/Q$
corrections in Drell-Yan) which can be summarized in general  terms as follows. Consider a typical `renormalon
integral'

\begin{equation}S(Q^2,N)=\int_{0}^{Q^2}{dk^2\over k^2} F(k^2/Q^2,N)
A_{{\cal S}}(k^2)\label{eq:ren-int},\end{equation} 
and introduce its (`RS invariant' \cite{Gru-Borel}) Borel
representation    by

\begin{equation}S(Q^2,N)=\int_0^\infty du\ \exp\left(-u\ln{Q^2 \over\Lambda^2}\right)\ B[S](u,N)
\label{eq:Borel-S}.\end{equation}
Using eq.(\ref{eq:Borel-coupling1}) one obtains the Borel transform of $S$ as

\begin{equation}B[S](u,N)=B[A_{{\cal S}}](u)\ {\tilde F}(u,N)
\label{eq:invBorel-S},\end{equation}
with

\begin{equation}{\tilde F}(u,N)=\int_0^1{dx\over x}\ F(x,N)\ \exp(-u\ln x)
\label{eq:F-tilde}.\end{equation}
The {\em factorized form} \cite{Gru-Borel,Beneke-Braun} of the expression should be noted. Suppose now
the Borel transform
$B[A_{{\cal S}}](u)$ of the effective coupling $A_{{\cal S}}$ has a zero at some position $u=u_0$, which is 
{\em not shared} by
$B[S](u,N)$. Then necessarily ${\tilde F}(u,N)$ in eq.(\ref{eq:invBorel-S}) must have a
pole\footnote{This argument is usually presented \cite{Beneke-Braun} as the existence of a zero in the Borel
transform $B[A_{{\cal S}}]$ causing the vanishing of the residue of a would-be renormalon contained in 
${\tilde F}(u,N)$. Thus this remark is a particular case of the observation in
\cite{Gru-Borel} that the renormalon residue is an {\em all-order} quantity, requiring the knowlegde of 
{\em all} the perturbative coefficients of the effective coupling $A_{{\cal S}}$.}
at $u=u_0$. If
$u_0>0$, this means an IR renormalon, and $F(x,N)$ contains an ${\cal O }(x^{u_0})$ contribution for
$x\rightarrow 0$. Since this renormalon is not present in $B[S](u,N)$, the standard IR renormalon philosophy
would conclude that no corresponding power correction is present. On the other hand, this power correction is
still expected in the IR finite coupling approach, where $A_{{\cal S}}(k^2)$ is assumed to have a finite IR
fixed point and the low energy part of the integral in eq.(\ref{eq:ren-int}) is well-defined. Then the
distribution function
$F(k^2/Q^2,N)$ can be expanded in powers of $k^2$ for $k^2\rightarrow 0$, yielding a non-vanishing power
correction for each term in its low energy expansion, parametrized by a low energy moment of the effective
coupling
$A_{{\cal S}}(k^2)$. Even if $B[S](u,N)$ is singular, rather than non-vanishing and finite, at $u=u_0$, there
is still a clash, since the singularity of ${\tilde F}(u,N)$ is necessarily stronger than that of $B[S](u,N)$
in presence of a zero of $B[A_{{\cal S}}](u)$; thus the IR finite coupling approach will predict a  {\em more
enhanced} power correction than indicated by the renormalon argument.
At this stage, there is no way to decide which of these two philosophies is correct without further
information. I end this section by giving two examples of this situation at large $N_f$.

i) DIS: there eq.(\ref{eq:Borel-coupling2}) implies that $B[A_{{\cal S}}](u)$ vanishes for any integer $u$,
$u\geq 3$.

ii) Drell-Yan: in this case the Sudakov effective coupling which occurs in the $Q^2$ derivative of the
Drell-Yan cross section is given by\footnote{The natural occurence of this combination of anomalous dimensions
has been pointed out independently from a different point of view by G.Sterman \cite{Sterman}.}

\begin{equation}8 C_F A_{{\cal S},DY}(k^2)= 2 A(a_s(k^2))+{dD_{DY}(a_s(k^2))\over d\ln
k^2}\label{eq:S-coupling-DY},\end{equation}
and the corresponding Borel transform at $N_f=\infty$ is \cite{Beneke-Braun} (dropping an inessential
 $\exp(c u)$ factor which corresponds simply to a change of scale):

\begin{equation}B[A_{{\cal S},DY}](u)={1\over \beta_0} {1\over \Gamma(1+u)} {\pi^{1/2}\over \Gamma(1/2-u)}
\label{eq:Borel-couplingDY},\end{equation}
which vanishes for positive half-integer $u$.

\noindent In both cases, the zeroes are absent from the corresponding Sudakov exponent, and the IR finite coupling
approach will lead to the prediction of extra power corrections in the exponent ($N^3/Q^6$, $N^4/Q^8$, etc...in
DIS, and $N/Q$, $N^3/Q^3$, etc...in Drell-Yan).

\section{Ambiguities in Sudakov resummation and exponentiation of constant terms}
To make progress, we have to understand better the origin of zeroes in the Borel transform 
$B[A_{{\cal S}}](u)$ of the Sudakov effective coupling. In this section, I shall show it is possible to
 make compatible the renormalon and IR finite coupling approaches. The main point which will be
developped is the following: there is an ambiguity in the choice
of the Sudakov distribution function (or, alternatively, in that of the Sudakov effective coupling),
which corresponds to the freedom to select an arbitrary set of 'constant terms' to be included in the Sudakov
exponent. One can eventually make use of this freedom to  reconcile the two conflicting approaches.

To see this, let us consider the specific case of DIS scattering eq.(\ref{eq:scale-viol}). Putting

\begin{equation}S(Q^2,N)=\int_{0}^1 dz{z^{N-1}-1 \over 1-z} A_{{\cal S}}[(1-z)Q^2]
\label{eq:S-DIS},\end{equation}
it is again convenient to write eq.(\ref{eq:S-DIS}) as a renormalon integral (eq.(\ref{eq:ren-int})), where
the Sudakov distribution function is now given by

\begin{equation}F(k^2/Q^2,N)=(1-k^2/Q^2)^{N-1}-1
\label{eq:F-DIS}.\end{equation}
I first note that order by order the perturbative series of $S(Q^2,N)$ contain both ${\cal O}(N^0)$ `constant
terms' and terms which vanish as $N\rightarrow\infty$, e.g.

\begin{equation}S(Q^2,N)=(\gamma_{01}L+\gamma_{00})a_s(Q^2)+(\gamma_{12}L^2+\gamma_{11}L+\gamma_{10})a_s^2(Q^2)
+...+{\cal O}(1/N)
\label{eq:S-DIS-series},\end{equation}
where $L\equiv \ln N$.

 A very useful simplification is achieved  by making use
of the following important scaling property: at large $N$, $F(k^2/Q^2,N)$ becomes a function of the variable
$\epsilon={N k^2\over Q^2}$ only\footnote{For Drell-Yan, the
corresponding scaling variable is $\epsilon_{DY}={N k\over Q}$.}, reflecting the relevance \cite{ABCMV} of
the `soft scale'
$Q^2/N$.
Indeed defining

\begin{equation}F(k^2/Q^2,N)\equiv G(\epsilon,N)=(1-\epsilon/N)^{N-1}-1
\label{eq:G-DIS},\end{equation}
and taking the $N\rightarrow \infty$ limit with {\em $\epsilon$  fixed} one gets a
finite result

\begin{equation} G(\epsilon,N)\rightarrow G(\epsilon,\infty)\equiv G(\epsilon)
\label{eq:G-inf1},\end{equation}
with

\begin{equation}  G(\epsilon)=\exp(-\epsilon)-1
\label{eq:G-inf2}.\end{equation}
 Let us now redefine the Sudakov exponent by using $
G(\epsilon)$ as the new distribution function, i.e. eq.(\ref{eq:ren-int}) is replaced by

\begin{equation}S_{stan}(Q^2,N)=\int_{0}^{Q^2}{dk^2\over k^2} G(N k^2/Q^2)
A_{{\cal S}}(k^2)\label{eq:ren-int-scaling},\end{equation}
where the index `$stan$' stands for `standard'. This step is legitimate since, order by order in perturbation
theory, $S_{stan}(Q^2,N)$ and
$S(Q^2,N)$ differ only by terms which vanish as $N\rightarrow\infty$, and thus share the {\em same} 
$\ln N$ and constant terms (eq.(\ref{eq:S-DIS-series})). Indeed, eq.(\ref{eq:ren-int-scaling}) can be written
identically as
 
\begin{equation}S_{stan}(Q^2,N)=\int_{0}^1 dz{\exp[-N(1-z)]-1 \over 1-z} A_{{\cal S}}[(1-z)Q^2]
\label{eq:S-DIS-alt},\end{equation}
and the equivalence between eq.(\ref{eq:S-DIS}) and eq.(\ref{eq:S-DIS-alt}) up to ${\cal O}(1/N)$ terms was
proved in \cite{CMW}. 

I now assume an ansatz of the form of eq.(\ref{eq:ren-int-scaling})

\begin{equation}S_{new}(Q^2,N)=\int_{0}^{Q^2}{dk^2\over k^2} G_{new}(N k^2/Q^2)
A_{{\cal S}}^{new}(k^2)\label{eq:ren-int-new},\end{equation}
and show that a {\em unique} solution for $G_{new}(N k^2/Q^2)$ and $A_{{\cal S}}^{new}(k^2)$ exists, under the
condition to reproduce all divergent $\ln N$ terms, together with an (a priori {\em arbitrary}) given set of
constant terms, i.e. 

\begin{equation}S_{new}(Q^2,N)=(\gamma_{01}L+\gamma_{00}^{new})a_s(Q^2)+(\gamma_{12}L^2+\gamma_{11}L+\gamma_{10}^{new})a_s^2(Q^2)
+...+{\cal O}(1/N)
\label{eq:Snew-DIS-series}.\end{equation}  
This statement can be checked order by order in perturbation theory.  As is well-known, the $\ln N$ structure in
eq.(\ref{eq:S-DIS-series}) or (\ref{eq:Snew-DIS-series})  arises from performing the integrals in
eq.(\ref{eq:ren-int-scaling}) or (\ref{eq:ren-int-new}) over the renormalization group $\ln(k^2/Q^2)$ logs which
show up when
$A_{{\cal S}}(k^2)$ is expanded in powers of $a_s(Q^2)$:

\begin{equation}A_{{\cal
S}}(k^2)=a_s(Q^2)+(-\beta_0\ln(k^2/Q^2)+{\cal A}_1)a_s^2(Q^2)+...\label{eq:AS-as}\end{equation} (where one should
set
$k^2=(1-z)Q^2$ if one uses instead eq.(\ref{eq:S-DIS}) or (\ref{eq:S-DIS-alt})). Using
eq.(\ref{eq:ren-int-scaling}) one thus gets

\begin{equation}S_{stan}(Q^2,N)=K_0(N) a_s(Q^2)+\left[-\beta_0 K_1(N)+{\cal A}_1 K_0(N)\right] a_s^2(Q^2)+...
\label{eq:S-DIS-series1},\end{equation}
with

\begin{eqnarray}K_p(N)&=&\int_{0}^{Q^2}{dk^2\over k^2} G(N k^2/Q^2)\ln^p(k^2/Q^2)\nonumber\\
&=&\int_{0}^{N}{d\epsilon\over \epsilon} G(\epsilon)\ln^p(\epsilon/N)
\label{eq:Kp}.\end{eqnarray}
Thus for large $N$ one gets (see also eq.(\ref{eq:K00}) and (\ref{eq:K11}))

\begin{equation}K_0(N)= \int_{0}^{N}{d\epsilon\over \epsilon} G(\epsilon)=c_{01}L+c_{00}+{\cal O}(1/N)
\label{eq:K0},\end{equation}
and

\begin{equation}K_1(N)= \int_{0}^{N}{d\epsilon\over \epsilon} G(\epsilon)\ln(\epsilon/N)=
c_{12}L^2+c_{11}L+c_{10}+{\cal O}(1/N)
\label{eq:K1},\end{equation}
where the $c_{ij}$'s are known. Let us now try to modify $G(\epsilon)\rightarrow G(\epsilon)+\Delta G(\epsilon)\equiv
G_{new}(\epsilon)$, in such a way the coefficients of all positive powers of  $\ln N$  remain {\em
unchanged}. I show that this cannot be achieved without changing simultaneously the Sudakov effective charge
$A_{{\cal S}}$. At
${\cal O}(a_s)$   one gets
$c_{00}\rightarrow c_{00}+\Delta c_{00}=c_{00}^{new}$, and one should require that

\begin{equation}\Delta c_{00}=\int_{0}^{\infty}{d\epsilon\over \epsilon}
\Delta G(\epsilon)<\infty
\label{eq:dc00}\end{equation}
in order to preserve the leading log coefficient $c_{01}$. However at ${\cal O}(a_s^2)$ one gets
$c_{11}\rightarrow c_{11}+\Delta c_{11}=c_{11}^{new}$ with

\begin{equation}\Delta c_{11}=-\int_{0}^{\infty}{d\epsilon\over \epsilon}
\Delta G(\epsilon)=-\Delta c_{00}\label{eq:dc11}.\end{equation}
Thus the single log term in $K_1(N)$ {\em must} be changed, and according to eq.(\ref{eq:S-DIS-series1}), this
implies a {\em correlated} change  ${\cal A}_1\rightarrow {\cal A}_1+\Delta {\cal A}_1={\cal A}_1^{new}$ if one
wants to preserve the subleading single
$\ln N$ term at ${\cal O}(a_s^2)$ in $S_{stan}$. Indeed, $\Delta c_{00}$ is already fixed (given $c_{00}$)
from the input
$c_{00}^{new}=\gamma_{00}^{new}$ (compare e.g. eq.(\ref{eq:S-DIS-series}) and (\ref{eq:S-DIS-series1})). Then, the
knowledge of $\Delta c_{11}$ fixes  $\Delta {\cal A}_1$ requiring
\begin{equation}-\beta_0\ \Delta c_{11}+\Delta {\cal A}_1\ c_{01}=0\label{eq:dc11-da1},\end{equation}
which fixes 
${\cal A}_1^{new}$, given ${\cal A}_1$.

Furthermore one gets
$c_{10}\rightarrow c_{10}+\Delta c_{10}=c_{10}^{new}$ with

\begin{equation}\Delta c_{10}=\int_{0}^{\infty}{d\epsilon\over \epsilon}
\Delta G(\epsilon)\ln\epsilon\label{eq:dc10}.\end{equation}
Thus $\Delta c_{10}$ is  fixed
(given
$c_{10}$ and
$A_1^{new}$) from the input
$-\beta_0\
c_{10}^{new}+A_1^{new}c_{00}^{new}=\gamma_{10}^{new}$ (compare again eq.(\ref{eq:S-DIS-series}) and
(\ref{eq:S-DIS-series1})). One thus sees that in this process $\Delta G(\epsilon)$ is determined {\em
uniquely} in principle by its input logarithmic moments $\Delta c_{i0}$ ($i: 0\rightarrow\infty$) (see
eq.(\ref{eq:dc00}) and (\ref {eq:dc10})), as can be easily checked in the first few orders of perturbation
theory. To determine
$\Delta G(\epsilon)$, we need however input {\em all} the $\gamma_{i0}^{new}$'s   and
use all orders of perturbation theory. This procedure is clearly not tractable, so we next turn to  large $N_f$
 where the problem can be solved in closed form.

\section{The asymptotic large $N$ Borel transform}
Before doing this, it is useful to first introduce
 a simplification appropriate to the large N limit, valid also at finite
$N_f$ beyond the single dressed gluon approximation. I observe that the perturbative series
eq.(\ref{eq:S-DIS-series})  of the Sudakov exponent eq.(\ref{eq:ren-int-scaling}) still contains ${\cal O}(1/N)$
terms at large $N$. It is useful to discard them, and find a Borel representation of the corresponding series 
$S_{as}(Q^2,N)$
  which contains only
$\ln N$ and ${\cal O}(N^0)$ terms, with all the ${\cal O}(1/N)$ terms in eq.(\ref{eq:S-DIS-series}) dropped.
The (RS invariant) Borel transform of eq.(\ref{eq:ren-int-scaling}) takes the factorized form (similarly to
eq.(\ref{eq:invBorel-S}) and (\ref{eq:F-tilde})) 

\begin{equation}B[S_{stan}](u,N)=B[A_{{\cal S}}](u) \exp(u\ln N) \int_0^N {d\epsilon\over\epsilon}
G(\epsilon) \exp(-u\ln\epsilon)\label{eq:B-stan}.\end{equation}
Taking straightforwardly the large $N$ limit one gets a tentative ansatz for the looked for Borel transform

\begin{equation}B[S_{as}](u,N)=B[A_{{\cal S}}](u) \exp(u\ln N) \int_0^{\infty}
{d\epsilon\over\epsilon} G(\epsilon) \exp(-u\ln\epsilon)\label{eq:B-as}.\end{equation}
Since the large $N_f$ calculation is usually performed in the `dispersive approach' \cite{BB, DMW} which
makes use of a `Minkowskian' coupling, it is convenient to use the equivalent `Minkowskian' representation of
$B[S_{as}](u,N)$, i.e. to introduce the Minkowskian counterpart of the Sudakov effective coupling,
together with a `Sudakov characteristic function' ${\cal G}(y)$ related to the Sudakov distribution
function $G(\epsilon)$ by the dispersion relation

\begin{equation}\dot{{\cal G}}(y)=y\int_0^{\infty} d\epsilon {G(\epsilon)\over (y+\epsilon)^2}
\label{eq:disp},\end{equation}
where $\dot{{\cal G}}=-y d{\cal G}/ dy$. It follows that\footnote{$B[A_{{\cal S}}](u)
 {\sin\pi u\over\pi u}$ is the Borel transform of the Minkowskian Sudakov  effective
coupling.}

\begin{equation}B[S_{as}](u,N)=B[A_{{\cal S}}](u) {\sin\pi u\over\pi u} \exp(u\ln N) \int_0^{\infty}
{dy\over y} \dot{{\cal G}}(y) \exp(-u\ln y)\label{eq:B-as-disp}.\end{equation}
Comparing eq.(\ref{eq:B-as}) with eq.(\ref{eq:B-as-disp}) yields  the standard relation between `Euclidean' and
`Minkowskian' Laplace transforms 
\begin{equation}\int_0^{\infty}
{d\epsilon\over\epsilon} G(\epsilon) \exp(-u\ln\epsilon)={\sin\pi u\over\pi u} \int_0^{\infty}
{dy\over y} \dot{{\cal G}}(y) \exp(-u\ln y)
\label{eq:BE-BD}.\end{equation}
However, eq.(\ref{eq:B-as}) and (\ref{eq:B-as-disp}) are not quite correct. For $u=0$ the
integrals on the right hand side are UV divergent, since $\dot{{\cal G}}(\infty)=G(\infty)=-1$ (see
eq.(\ref{eq:G-inf2})). One actually needs to `renormalize' these `bare' Borel transforms, i.e. to add a
subtraction term. One can show that the correct answer, in the case of the Euclidean representation, is 

\begin{equation}B[S_{as}](u,N)=B[A_{{\cal S}}](u) \left[ \exp(u\ln N)
\int_0^{\infty}
{d\epsilon\over\epsilon} G(\epsilon)
\exp(-u\ln\epsilon)-{G(\infty)\over u}\right]\label{eq:B-as-euclid-ren},\end{equation}
whereas for the Minkowskian representation (as follows from eq.(\ref{eq:BE-BD}))

\begin{equation}B[S_{as}](u,N)=B[A_{{\cal S}}](u) {\sin\pi u\over\pi u}\left[ \exp(u\ln N)
\int_0^{\infty} {dy\over y} \dot{{\cal G}}(y) \exp(-u\ln
y)-{\pi\dot{{\cal G}}(\infty)\over \sin\pi u}\right]\label{eq:B-as-disp-ren}.\end{equation}
The corresponding `renormalon integral'  representation of $S_{as}(Q^2,N)$ analogous to
eq.(\ref{eq:ren-int-scaling}) is 

\begin{eqnarray}S_{as}(Q^2,N)&=&\int_{0}^{\infty}{dk^2\over k^2} G(N k^2/Q^2)
A_{{\cal S}}(k^2)-G(\infty)\int_{Q^2}^{\infty}{dk^2\over k^2} A_{{\cal S}}(k^2)\label{eq:ren-int-scaling-as}\\
&\equiv&\int_{0}^{Q^2}{dk^2\over k^2} G(N k^2/Q^2)
A_{{\cal S}}(k^2)+\int_{Q^2}^{\infty}{dk^2\over k^2}\left[G(N k^2/Q^2)-G(\infty)\right] A_{{\cal
S}}(k^2)\nonumber,\end{eqnarray} where the second integral on the right hand side on the first line provides
the necessary subtraction term (the first integral being UV divergent).

\section{Exponentiation of constant terms at large $N_f$: the single dressed gluon result}

\subsection{DIS case}
Specializing now to large $N_f$, the result for the Borel transform of $d\ln F_2(Q^2,N)/d\ln Q^2\vert _{SDG}$ at
{\em finite}
$N$ can be given
 in the  `massive gluon' or `single dressed gluon' (SDG)   dispersive Minkowskian formalism
\cite{BB,DMW} as (the V-scheme is assumed)

\begin{equation}B[d\ln F_2(Q^2,N)/d\ln Q^2]_{SDG}(u,N)=4 C_F{1\over\beta_0} {\sin\pi u\over\pi u}
\int_0^{\infty} {dx\over x} \ddot{{\cal F}}_{SDG}(x,N) \exp(-u\ln x)\label{eq:B-SDG},\end{equation}
where the `characteristic function' ${\cal F}_{SDG}(x,N)$ has been computed in \cite{DMW} 
($x=\lambda^2/Q^2$ where $\lambda$ is the `gluon mass'). I have checked that here too a similar scaling property
holds at large
$N$, namely, putting
${\cal G}_{SDG}(y,N)\equiv{\cal F}_{SDG}(x,N)$ with $y\equiv N x=N \lambda^2/Q^2$, one gets for
$N\rightarrow\infty$ at fixed $y$

\begin{equation} \ddot{{\cal G}}_{SDG}(y,N)\rightarrow \ddot{{\cal G}}_{SDG}(y,\infty)\equiv
\ddot{{\cal G}}_{SDG}(y)\label{eq:G-disp-inf},\end{equation}
and one can derive from the result of \cite{DMW} in the DIS case that

\begin{equation} \ddot{{\cal G}}_{SDG}(y)=-1+ \exp(-y)-{1\over 2}\ y\ \exp(-y)-{1\over 2}\ y\
\Gamma(0,y)+{1\over 2}\ y^2\
\Gamma(0,y)
\label{eq:Gdot-SDG}.\end{equation}
Thus for $N\rightarrow\infty$

\begin{equation}B[d\ln F_2(Q^2,N)/d\ln Q^2]_{SDG}^{as}(u,N)=4 C_F{1\over\beta_0} {\sin\pi u\over\pi u} \exp(u\ln N)
\int_0^{\infty} {dy\over y} \ddot{{\cal G}}_{SDG}(y) \exp(-u\ln y)\label{eq:B-SDG-as-disp},\end{equation}
which  has also to be subtracted

\begin{eqnarray}B[d\ln F_2(Q^2,N)/d\ln Q^2]_{SDG}^{as}(u,N)=4 C_F{1\over\beta_0} {\sin\pi u\over\pi u}\nonumber\\
\times\left[
\exp(u\ln N)
\int_0^{\infty} {dy\over y} \ddot{{\cal G}}_{SDG}(y) \exp(-u\ln y)
+{\Gamma_{SDG}(u)\over u}\right] \label{eq:B-SDG-as-disp-ren},\end{eqnarray}
where $\Gamma_{SDG}(0)=1$ is finite. One can show that

\begin{equation}{\Gamma_{SDG}(u)\over u}=\int_0^{\infty} {dy\over y} \left[\ddot{\nu}(y)+1\right]
\exp(-u\ln y)\label{eq:Gamma-SDG1},\end{equation}
where $\nu (y)$ is the ({\em universal}) virtual contribution \cite{DMW}  to
the massive gluon characteristic function for space-like processes\footnote{The normalization is half that of
\cite{DMW}.} 

\begin{equation}\nu (y)=-\int_0^1 dz {(1-z)^2\over z-y}\ln{z\over y}\label{eq:nu}.\end{equation}
It follows that

\begin{equation}\Gamma_{SDG}(u)=\left({\pi u\over\sin\pi
u}\right)^2{1\over (1-u)(1-u/2)}\label{eq:Gamma-SDG2}.\end{equation}
The above mentionned fact that $\nu (y)$ is universal suggests that $\Gamma_{SDG}(u)$ might also be universal for
all space-like processes.

I note that the constant terms of the SDG
result are obtained by setting
$N=1$ in eq.(\ref{eq:B-SDG-as-disp-ren}). Thus, if one wants to select as input an {\em arbitrary} subset of
constant terms to be included into a {\em new} asymptotic Sudakov exponent $S_{\tilde {as}}(Q^2,N)$, one should
simply change the subtraction function
$\Gamma_{SDG}(u)$ in eq.(\ref {eq:B-SDG-as-disp-ren}), namely define a new input $B[d\ln F_2(Q^2,N)/d\ln
Q^2]_{SDG}^{\tilde {as}}$ by

\begin{eqnarray}B[d\ln F_2(Q^2,N)/d\ln Q^2]_{SDG}^{\tilde {as}}(u,N)=4 C_F{1\over\beta_0} {\sin\pi u\over\pi
u}\nonumber\\
\times\left[
\exp(u\ln N)
\int_0^{\infty} {dy\over y} \ddot{{\cal G}}_{SDG}(y) \exp(-u\ln y)+{\Gamma_{SDG}^{new}(u)\over u}\right]
\label{eq:B-SDG-asnew-disp-ren},\end{eqnarray}
where $\Gamma_{SDG}^{new}(u)$ (still with $\Gamma_{SDG}^{new}(0)=1$) takes into account the new selected set of
constant terms to be included together with the $\ln N$'s into the new Sudakov exponent. The latter (output)
should thus be given by (see eq.(\ref{eq:B-as-disp-ren}))

\begin{equation}B[S_{\tilde {as}}](u,N)=B[A_{{\cal S}}^{new}](u) {\sin\pi u\over\pi u}\left[ \exp(u\ln N)
\int_0^{\infty} {dy\over y} \dot{{\cal G}}_{new}(y) \exp(-u\ln
y)-{\pi\dot{{\cal G}}(\infty)\over \sin\pi u}\right]\label{eq:B-astlid-disp-ren},\end{equation}
where $\dot{{\cal G}}(\infty)=-1$ does not change (it determines the leading logs). Since there are no ${\cal
O}(1/N)$ terms, one can now identify

\begin{equation}4 C_F B[S_{\tilde {as}}](u,N)\equiv B[d\ln
F_2(Q^2,N)/d\ln Q^2]_{SDG}^{\tilde {as}}(u,N)\label{eq:BS-BF},\end{equation}
which yields the {\em two} master equations

\begin{equation}B[A_{{\cal S}}^{new}](u) \int_0^{\infty} {dy\over y}
 \dot{{\cal G}}_{new}(y) \exp(-u\ln y)={1\over\beta_0}\int_0^{\infty}{dy\over y} \ddot{{\cal G}}_{SDG}(y)
\exp(-u\ln y)\label{eq:A-G-SDG},\end{equation}
and
\begin{eqnarray}B[A_{{\cal S}}^{new}](u)&=&{1\over\beta_0}\left({\sin\pi u\over\pi u}\right)
\left(-{\Gamma_{SDG}^{new}(u)\over
\dot{{\cal G}}(\infty)}\right)\nonumber\\
&=&{1\over\beta_0}\left({\sin\pi u\over\pi u}\right) \Gamma_{SDG}^{new}(u)\label{eq:A-G}.\end{eqnarray}
It also follows from eq.(\ref{eq:Gdot-SDG}) that

\begin{equation}\int_0^{\infty}{dy\over y} \ddot{{\cal G}}_{SDG}(y) \exp(-u\ln
y)=\Gamma(-u){1\over 2}\left({1\over
1-u}+{1\over 1-u/2}\right)\label{eq:Mellin-SDG}.\end{equation}
Eq.(\ref{eq:A-G-SDG}) and (\ref{eq:A-G}) allow to determine {\em both} the Sudakov characteristic function
${\cal G}_{new}(y)$ and the Borel transform of the associated Sudakov effective coupling $B[A_{{\cal
S}}^{new}](u)$ corresponding to a given input set of exponentiated constant terms which fix the subtraction
function  $\Gamma_{SDG}^{new}(u)$. It is interesting that $B[A_{{\cal S}}^{new}](u)$ is given {\em entirely}
by the subtraction term. Once  ${\cal G}_{new}(y)$ is determined, one can derive the corresponding
Sudakov distribution function $G_{new}(\epsilon)$ using eq.(\ref{eq:disp}) ($G_{new}(\epsilon)$ is essentially
the time-like discontinuity of ${\cal G}_{new}(y)$), thus proving at large
$N_f$ the statement made below eq.(\ref{eq:ren-int-new}). 

In particular, one can easily obtain the result in the two extreme cases where {\em all} constant terms are
included in the Sudakov exponent (the input is then given by eq.(\ref{eq:B-SDG-as-disp-ren})), and that where
{\em none} are. But these two cases are not interesting for the IR finite coupling approach, since one finds
that $B[A_{{\cal S}}^{new}](u)$ then contains renormalons. In particular, if one tries to include all constant
terms one gets using eq.(\ref{eq:Gamma-SDG2})

\begin{equation}B[A_{{\cal S}}^{all}](u)={1\over\beta_0}\left({\pi u\over \sin\pi u}\right){1\over
(1-u)(1-u/2)}\label{eq:AS-all}.\end{equation}
Despite the presence of IR renormalon poles, this result might be of interest since this coupling is expected to
be universal for all space-like processes according to the remark made below eq.(\ref{eq:Gamma-SDG2}).
However, the corresponding result for the Sudakov characteristic function

\begin{equation}\int_0^{\infty} {dy\over y} \dot{{\cal G}}_{all}(y) \exp(-u\ln y)=-{1\over u}{1\over
\Gamma(1+u)}\left(1-{3 u\over 4}\right)
\label{eq:GS-all}\end{equation}
shows that the latter actually {\em does not} exist, the inverse gamma function having no Mellin representation!

On the other hand, if {\em no} constant terms are included in the Sudakov exponent, one finds that the
corresponding Sudakov distribution function
$G_{logs}(\epsilon)$ is cut-off in the infrared

\begin{equation}G_{logs}(\epsilon)=-\theta(\epsilon-1)\label{eq:GS-none},\end{equation} 
so that IR renormalons can show up only
through the divergent expansion of the Sudakov effective coupling itself. Indeed one gets 

\begin{equation}B[A_{{\cal S}}^{logs}](u)={1\over\beta_0}{1\over
\Gamma(1+u)}{1\over 2}\left({1\over
1-u}+{1\over 1-u/2}\right)\label{eq:AS-none},\end{equation}
which has IR renormalon poles at $u=1,2$.

More interesting is the simplest possible ansatz, namely $\Gamma_{SDG}^{simple}(u)\equiv 1$, which gives
$B[A_{{\cal S}}^{simple}](u)={1\over\beta_0}{\sin\pi u\over\pi u}$ and corresponds to $A_{{\cal S}}^{simple}(k^2)$
(eq.(\ref{eq:A-simple})), which is already IR finite at the {\em perturbative} level. However, the first two
zeroes at
$u=1$ and
$u=2$  lead to two  {\em log-enhanced} power corrections at large $N$ in the IR finite coupling
framework: the right hand side of eq.(\ref{eq:Mellin-SDG}) has {\em double} IR renormalon poles at
$u=1,2$, resulting in ${\cal O}(y\ln^2 y)$ and ${\cal O}(y^2\ln^2 y)$ terms at small $y$ in $\dot{{\cal
G}}_{simple}(y)$. In fact   the Sudakov distribution function is now given by

\begin{equation}G_{simple}(\epsilon)=\ddot{{\cal G}}_{SDG}(\epsilon)\label{eq:G-simple},\end{equation}
and eq.(\ref{eq:Gdot-SDG}) indeed yields as $\epsilon\rightarrow 0$

\begin{equation}\ddot{{\cal G}}_{SDG}(\epsilon)\simeq {1\over
2}[\epsilon(\ln\epsilon+\gamma_E-3)+\epsilon^2(-\ln\epsilon-\gamma_E+1)+{5\over 12}\epsilon^3-{1\over
18}\epsilon^4 +...]
\label{eq:G-SDG-IR}.\end{equation}
There is thus a discrepancy with the IR renormalon expectation, in agreement with the argument of section 4, and
the question  arises whether such an ansatz would violate the OPE at large $N$ already at the level of the two
leading ($N/Q^2$ and
$N^2/Q^4$) power corrections. Given the well-known
\cite{Sterman-OPE, GK} intricacies of the latter, the question is perhaps not yet settled.

Applying these results to the standard case, where
$G(\epsilon)$ is known (eq.(\ref{eq:G-inf2})) and fixes $\dot{{\cal G}}(y)$ from eq.(\ref{eq:disp}), and using
eq.(\ref{eq:A-G-SDG}), one can also rederive  the result  eq.(\ref{eq:Borel-coupling2}), where the first two zeroes
are  cancelled. Here the discrepancy with the IR renormalon prediction concerns only the higher order power
corrections ($N^3/Q^6$, $N^4/Q^8$,...), which are expected \cite{GK, Gardi-Roberts} to be absent at large $N$ in
the exponent.

\subsection{Drell-Yan case: emergence of an IR finite {\em perturbative} effective  coupling}
The analogue of eq.(\ref{eq:scale-viol} ) for the scaling violation of the short distance Drell-Yan cross-section
is

\begin{equation}{d\ln \sigma_{DY}(Q^2,N)\over d\ln Q^2}=4 C_F\left( H_{DY}(Q^2)+S_{DY}(Q^2,N)\right)+{\cal
O}(1/N)  
\label{eq:scale-viol-DY},\end{equation}
with

\begin{equation}S_{DY}(Q^2,N)=\int_{0}^1 dz\ 2{z^{N-1}-1 \over
1-z} A_{{\cal S},DY}[(1-z)^2 Q^2]  
\label{eq:S-DY}.\end{equation}
$S_{DY}(Q^2,N)$ can again be written as a renormalon integral

\begin{equation}S_{DY}(Q^2,N)=\int_{0}^{Q^2}{dk^2\over k^2} F_{DY}(k^2/Q^2,N)
A_{{\cal S},DY}(k^2)\label{eq:ren-int-DY},\end{equation}
where $A_{{\cal S},DY}(k^2)$ is given in eq.(\ref{eq:S-coupling-DY}), and the Sudakov distribution function

\begin{equation}F_{DY}(k^2/Q^2,N)=(1-k/Q)^{N-1}-1
\label{eq:F-DY},\end{equation}
 involves {\em both} \cite{GS} even and odd powers of $k$ at small $k$. Taking the scaling limit
$N\rightarrow\infty$ with $\epsilon_{DY}=Nk/Q$ fixed one thus get the analogue of eq.(\ref{eq:G-inf2})

\begin{equation}  G_{DY}(\epsilon_{DY})=\exp(-\epsilon_{DY})-1
\label{eq:G-DY}.\end{equation}
Now the work of \cite{Vogelsang} for eikonal cross sections suggests to exponentiate a new set of constant terms
(for any
$N_f$) using

\begin{equation}S_{DY}^{new}(Q^2,N)=\int_{0}^{Q^2}{dk^2\over k^2} G_{DY}^{new}(N k/Q)
A_{{\cal S},DY}^{new}(k^2)\label{eq:ren-int-new-DY},\end{equation}
with (we deal with the log-derivative of the Drell-Yan cros-section)

\begin{equation}G_{DY}^{new}(\epsilon_{DY})=2 {d\over d\ln Q^2}\left[K_0(2
N k/Q)+\ln(N k/Q)+\gamma_E\right]\label{eq:G-new-DY1},\end{equation}
i.e.
 
\begin{equation}G_{DY}^{new}(\epsilon_{DY})=-\left[1+x{dK_0\over dx}(x=2
\epsilon_{DY})\right]\label{eq:G-new-DY},\end{equation}
where $K_0(x)$ is the modified Bessel function of the second kind.\footnote{
$G_{DY}^{new}(\epsilon_{DY})\rightarrow -1$ for $\epsilon_{DY}\rightarrow \infty$, consistently with the large
$\epsilon_{DY}$ limit of eq.(\ref{eq:G-DY}).} 
But in the Drell-Yan case the analogue of
eq.(\ref{eq:B-as}) is

\begin{equation}B[S_{DY,as}](u,N)=B[A_{{\cal S},DY}](u) \exp(2 u\ln N) \int_0^{\infty}
2{d\epsilon\over\epsilon} G_{DY}(\epsilon) \exp(-2 u\ln\epsilon)\label{eq:B-as-DY},\end{equation}
and at large $N_f$  we get, instead\footnote{Recall also eq.(\ref{eq:BE-BD}).} of eq.(\ref{eq:A-G-SDG})

\begin{equation}B[A_{{\cal S},DY}^{new}](u) \int_0^{\infty}
2{d\epsilon\over\epsilon} G_{DY}^{new}(\epsilon) \exp(-2 u\ln\epsilon)={1\over\beta_0}{\sin\pi u\over\pi
u}\int_0^{\infty}{dy\over y} \ddot{{\cal G}}_{SDG,DY}(y)
\exp(-u\ln y)\label{eq:A-G-SDG-DY}.\end{equation}
On the other hand from the large $N_f$ calculation of \cite{Beneke-Braun} one gets the $N$-dependent part of the
large $N$ Borel transform

\begin{equation}{\sin\pi u\over\pi
u}\int_0^{\infty}{dy\over y} \ddot{{\cal G}}_{SDG,DY}(y)
\exp(-u\ln y)=-{1\over u}{\Gamma(1-u)\over \Gamma(1+u)}\label{eq:B-SDG-DY},\end{equation}
whereas  eq.(\ref{eq:G-new-DY}) yields

\begin{equation}\int_0^{\infty}
2{d\epsilon\over\epsilon} G_{DY}^{new}(\epsilon) \exp(-2 u\ln\epsilon)=-u
[\Gamma(-u)]^2\label{eq:B-K0}.\end{equation} 
From eq.(\ref{eq:A-G-SDG-DY}) one thus derive the large $N_f$ result

\begin{equation}B[A_{{\cal S},DY}^{new}](u)={1\over\beta_0}{1\over
\Gamma(1+u)\Gamma(1-u)}\equiv{1\over\beta_0}{\sin\pi u\over\pi u}
\label{eq:A-new-DY},\end{equation} 
i.e. $A_{{\cal S},DY}^{new}(k^2)$ is nothing but the effective coupling $A_{{\cal S}}^{simple}(k^2)$ of
eq.(\ref{eq:A-simple}). The latter,
as we have seen, does have an IR finite fixed point, and no non-perturbative modification is a priori necessary in
this case! As expected from the general discussion in section 4, the zeroes in $B[A_{{\cal S},DY}^{new}](u)$ lead
to large $N$ logarithmically enhanced power corrections in the IR finite coupling framework, at variance with the
IR renormalon expectation. Indeed the new Sudakov distribution function has logarithmically enhanced contributions
for $\epsilon_{DY}\rightarrow 0$

\begin{equation}G_{DY}^{new}(\epsilon_{DY})=\epsilon_{DY}^2(2\ln\epsilon_{DY}+2\gamma_E-1)+
\epsilon_{DY}^4(\ln\epsilon_{DY}+\gamma_E-{5\over 4})+{\cal
O}(\epsilon_{DY}^6\ln\epsilon_{DY})\label{eq:G-new-DY-IR}.\end{equation}
The fact that this new Sudakov distribution function implies a new Sudakov effective coupling is of course one of
the main point of the present paper.

\section{Reconciling the IR renormalon and the IR finite coupling approaches: a large $N_f$ ansatz for a
`universal' Sudakov effective coupling}  I next turn to the question raised in the beginning of  section 5 how to
reconcile the IR renormalon and IR
finite coupling approaches to power corrections.  Actually, I should first stress it is not yet clear whether the
two approaches should be necessarily reconciled. For instance, in the DIS case, it could be that the OPE at large
$N$ is consistent with the existence of two leading log-enhanced power corrections, as predicted  by the 
$A_{{\cal S}}^{simple}(k^2)$ ansatz (eq.(\ref{eq:A-simple})), or with the existence of higher order power
corrections ($N^3/Q^6$, $N^4/Q^8$,...) at large $N$ {\em in the exponent} (see section 7). If this turns out to
be the case,  the IR finite coupling approach would be consistent
with the OPE (and at odds with the IR renormalon prediction) with the `simple' ansatz of eq.(\ref{eq:A-simple}), or
eventually with the standard result of eq.(\ref{eq:Borel-coupling2}). Similarly, the `simple' ansatz of
eq.(\ref{eq:A-simple}) might be the correct one in the Drell-Yan case, arising from a  `natural'
exponentiation of some ${\cal O}(N^0)$ terms, despite it also contradicts the standard IR renormalon expectation.
It is interesting to note at this point  that eq.(\ref{eq:A-G}) indicates that zeroes in $B[A_{{\cal
S}}^{new}](u)$ arise from {\em two} distinct sources: either the `universal' $\sin\pi u/\pi u$ factor ({\em
simple} zeroes at integer
$u$ can come only from there), or the `arbitrary' $ \Gamma_{SDG}^{new}(u)$ subtraction term (zeroes at
half-integer $u$ can come only from there).
The previous discussion  suggests that  zeroes coming from the `universal'
$\sin\pi u/\pi u$ factor need {\em not} be necessarily removed in the IR finite coupling approach, at the
difference of the more `artificial' zeroes (such as those occuring in the standard Drell-Yan case
eq.(\ref{eq:Borel-couplingDY})) coming from the subtraction term.

Notwithstanding the above remarks, I shall adopt in this section the attitude that the IR renormalon and IR
finite coupling approaches to power corrections should {\em always} be made consistent with one another. For this
purpose, one  must  remove {\em all}
\footnote{Except  those eventually also present in
$\int_0^{\infty}{dy\over y}
\ddot{{\cal G}}_{SDG}(y)
\exp(-u\ln y)$, such as the one at $u=4/3$ (see eq.(\ref{eq:Mellin-SDG})).} zeroes from
$B[A_{{\cal S}}^{new}](u)$. The {\em mathematically simplest} \footnote{Apart from the obvious choice
$\Gamma_{SDG}^{new}(u)=\pi u/\sin\pi u$, where the Sudakov effective coupling is just the one-loop coupling,
and is thus of little interest (having a Landau pole) for the IR finite coupling approach.}  ansatz suggested
by eq.(\ref{eq:A-G}) is to choose

\begin{equation}\Gamma_{SDG}^{new}(u)=\Gamma(1-u)\label{eq:Gamma}\end{equation}
which yields

\begin{equation}B[A_{{\cal S}}^{new}](u)={1\over\beta_0}\left({\sin\pi u\over\pi u}\right)
\Gamma(1-u)={1\over\beta_0}{1\over \Gamma(1+u)}\label{eq:BA-QCD}.\end{equation}
It is interesting to compare this ansatz with the result obtained by eliminating the half-integer zeroes from the
Drell-Yan standard result eq.(\ref{eq:Borel-couplingDY}). There the simplest ansatz is to  define

\begin{equation}B[A_{{\cal S},DY}^{new}](u)= {\Gamma(1/2-u)\over
\pi^{1/2}}B[A_{{\cal S},DY}](u)
\label{eq:Borel-couplingDYnew},\end{equation}
which yields again the `universal' ansatz eq.(\ref{eq:BA-QCD}).

Given an ansatz for $B[A_{{\cal S}}^{new}](u)$, eq.(\ref{eq:A-G-SDG}) then determines $\dot{{\cal G}}_{new}(y)$,
hence the Sudakov distribution function $G_{new}(\epsilon)$. In particular, in the DIS case one gets assuming the
ansatz eq.(\ref{eq:BA-QCD}) 

\begin{eqnarray}G_{new}(\epsilon)&=&-{\epsilon(1+\epsilon)\over 2}\ \  0\leq\epsilon\leq 1\nonumber\\
&=&-1\ \ \ \ \ \ \ \ \ \ \ \  \epsilon\geq 1\label{eq:Gnew},\end{eqnarray}
where $\epsilon=N k^2/Q^2$. Thus all power corrections beyond the two leading ones are indeed absent from the
Sudakov exponent, in agreement with the renormalon argument.

\noindent\underline{Infrared behavior:} although I have not been able to find a closed form analytic expression
for the  Sudakov effective coupling corresponding to eq.(\ref{eq:BA-QCD})

\begin{equation}A_{{\cal S}}^{new}(k^2)={1\over\beta_0}\int_0^\infty du\
\exp\left(-u\ln{k^2 \over\Lambda^2}\right)\
{1\over \Gamma(1+u)} \label{eq:QCD-coupling},\end{equation}
there is very strong numerical evidence that it blows up very fast (but remains positive) for $k^2\rightarrow 0$:

\begin{equation}A_{{\cal S}}^{new}(k^2)\simeq {1\over \beta_0}\exp\left({\Lambda^2\over k^2}\right)
\label{eq:couplingQCDIR}.\end{equation}
Assuming that eq.(\ref{eq:couplingQCDIR}) is correct, one can again speculate  that there exists a
non-perturbative modification $\delta A_{{\cal S},NP}(k^2)$ of the coupling at very small momenta which will
eventually turn the too strong  IR singularity  into a  non-perturbative, but 
finite IR fixed point.  A simple ansatz for $\delta
A_{{\cal S},NP}(k^2)$, localized in the IR region, can be constructed in analogy with
eq.(\ref{eq:deltacoupling}). Expanding the righthand side of eq.(\ref{eq:couplingQCDIR}) for {\em large} $k^2$

\begin{equation}{1\over \beta_0}\exp\left({\Lambda^2\over k^2}\right)={1\over \beta_0}\left(1+{\Lambda^2\over
k^2}+{1\over 2}{\Lambda^4\over k^4}+{1\over 3!}{\Lambda^6\over k^6}+...\right)
\label{eq:expand},\end{equation}
suggests to subtract term by term using

\begin{equation}\delta A_{{\cal S},NP}^{new}(k^2)=-{1\over \beta_0}\left[{\Lambda^2\over
k^2}\exp\left(-{k^2\over \Lambda^2}\right)+{1\over 2}{\Lambda^4\over k^4}\exp\left(-{k^4\over
\Lambda^4}\right)+{1\over 3!}{\Lambda^6\over k^6}\exp\left(-{k^6\over
\Lambda^6}\right)+...\right]
\label{eq:expand}.\end{equation}
Then for $k^2\rightarrow 0$ one gets

\begin{equation}\delta A_{{\cal S},NP}^{new}(k^2)\simeq -{1\over \beta_0}\left({\Lambda^2\over
k^2}+{1\over 2}{\Lambda^4\over k^4}+{1\over 3!}{\Lambda^6\over k^6}+...\right)+ {1\over \beta_0}\left(1+{1\over
2}+{1\over 3!}+...\right)+{\cal O}(k^2/\Lambda^2)
\label{eq:expanddelta}.\end{equation}
Thus defining

\begin{equation}A_{{\cal S},NP}^{new}(k^2)=A_{{\cal S}}^{new}(k^2)+\delta A_{{\cal S},NP}^{new}(k^2)
\label{eq:npcouplingQCD}\end{equation}
 yields indeed for $k^2\rightarrow 0$  a finite and positive IR fixed point
\begin{equation}A_{{\cal S},NP}^{new}(k^2)\simeq{1+e\over\beta_0}+{\cal
O}(k^2)\label{eq:npcouplingQCDIR}.\end{equation}

\section{Resummation of the `left-over' constant terms at large $N_f$, and a strategy for phenomenology at finite
$N_f$} For a given choice of the exponentiated constant terms to be included in the Sudakov exponent
$S_{new}(Q^2,N)$ (eq.(\ref{eq:ren-int-new})), there remains corresponding `left-over' constant terms $H_{new}(Q^2)$
(see also eq.(\ref{eq:scale-viol}))

\begin{equation}{d\ln F_2(Q^2,N)\over d\ln Q^2}=4 C_F\left( H_{new}(Q^2)+S_{new}(Q^2,N)\right)+{\cal O}(1/N)  
\label{eq:scale-viol-new},\end{equation}
with

\begin{equation}H_{new}(Q^2)= \Delta\gamma_{00}^{new} a_s(Q^2)+\Delta\gamma_{10}^{new}
a_s^2(Q^2)+...\label{eq:H-new},\end{equation}
such that

\begin{equation}\gamma_{i0}^{new}+ \Delta\gamma_{i0}^{new}=g_i\label{eq:g},\end{equation}
where the $g_i$'s are the full constant terms at large $N$

\begin{equation}{d\ln F_2(Q^2,N)\over d\ln
Q^2}=4 C_F\left[(\gamma_{01}L+g_{0})a_s(Q^2)+(\gamma_{12}L^2+\gamma_{11}L+g_{1})a_s^2(Q^2) +...\right]
+{\cal O}(1/N)
\label{eq:DIS-series}.\end{equation}
From eq.(\ref{eq:B-SDG-as-disp-ren}) and (\ref{eq:B-SDG-asnew-disp-ren}) it follows that the Borel transform of
$H_{new}(Q^2)$ at large $N_f$ is given by (in the V-scheme)

\begin{equation}B[H_{new}(Q^2)](u)={1\over \beta_0}\left({\sin\pi u\over\pi u}\right)
{\Gamma_{SDG}(u)-\Gamma_{SDG}^{new}(u)\over u}\label{eq:BH1}.\end{equation}
It is natural to try resumming the $H_{new}(Q^2)$ series with a renormalon integral representation

\begin{equation}H_{new}(Q^2)=\int_{0}^{\infty}{dk^2\over k^2} G_{0,new}(k^2/Q^2)
A_{{\cal S}}^{new}(k^2)\label{eq:ren-int-Hnew},\end{equation}
using the {\em same} effective coupling $A_{{\cal S}}^{new}(k^2)$ as in the Sudakov exponent, which {\em defines}
the `left-over constants'  distribution function $G_{0,new}(k^2/Q^2)$. Eq.(\ref{eq:ren-int-Hnew}) yields a
representation for the Borel transform

\begin{equation}B[H_{new}(Q^2)](u)=B[A_{{\cal S}}^{new}](u) \int_0^{\infty}
{dx\over x} G_{0,new}(x) \exp(-u\ln x)\label{eq:BH2}.\end{equation}
Comparing eq.(\ref{eq:BH1}) with eq.(\ref{eq:BH2})  allows to determine $G_{0,new}(k^2/Q^2)$. For instance,
in the case of the `new Sudakov effective coupling' eq.(\ref{eq:BA-QCD}), one finds that ($x=k^2/Q^2$)

\begin{equation}G_{0,new}(x)=2 x\left[(2x-1)\exp(1/x)\Gamma(0,1/x)-2x(\ln x-\gamma_E)-(\ln
x-\gamma_E+2)\right]-\theta(x-1)
\label{eq:G0},\end{equation}
and one can check that $G_{0,new}(x)={\cal O}(\ln x/x)$ for $x\rightarrow\infty$, while for $x\rightarrow 0$

\begin{equation}G_{0,new}(x)\simeq -2 x(\ln x-\gamma_E+2)\label{eq:G0-IR}.\end{equation}
Eq.(\ref{eq:G0-IR}) predicts in the IR finite coupling approach a  log-enhanced ${\cal O}(\ln Q^2/Q^2)$
$N$-independent power correction arising from the left-over constant terms. This is consistent with the IR
renormalon expectation as can be checked by setting
$N=1$ in eq.(\ref{eq:B-SDG-as-disp-ren}).

\noindent\underline{Application to phenomenology at finite $N_f$}: I suggest to use the {\em same} large $N_f$
Sudakov distribution function $G_{new}(\epsilon)$ and `left-over constants' distribution function $G_{0,new}(x)$ at
{\em finite} $N_f$, while adjusting the corresponding Sudakov effective coupling $A_{{\cal S}}^{new}(k^2)$ to the
finite $N_f$ situation. For instance, assuming the  ansatz eq.(\ref{eq:QCD-coupling}) is the correct one at
large $N_f$, one would use the finite
$N_f$ resummation formula 

\begin{eqnarray}{d\ln F_2(Q^2,N)\over d\ln Q^2}&=&4 C_F\int_{0}^{Q^2}{dk^2\over k^2} G_{new}(N k^2/Q^2)
A_{{\cal S}}^{new}(k^2)\label{eq:scale-viol-phen}\\
&+&4 C_F \int_{0}^{\infty}{dk^2\over k^2} G_{0,new}(k^2/Q^2)
A_{{\cal S}}^{QCD}(k^2)+4 C_F\ \Delta H_{new}(Q^2)+{\cal O}(1/N)\nonumber,\end{eqnarray}
where the ({\em $N_f$-independent}) distribution functions $G_{new}(N k^2/Q^2)$ and $G_{0,new}(k^2/Q^2)$ are the
{\em same} as in eq.(\ref{eq:Gnew}) and (\ref{eq:G0}), while $A_{{\cal S}}^{new}(k^2)$ can be determined order by
order in perturbation theory at finite
$N_f$ in a standard way,  matching  eq.(\ref{eq:scale-viol-phen}) with the perturbative expansion of the left hand
side (only the $N$-dependent terms, i.e. the first line, are necessary for this purpose). I note that at finite
$N_f$ there may still remain `relic' constant terms contained in the function $\Delta H_{new}(Q^2)$, not accounted
for by the two renormalon integrals on the right hand side of eq.(\ref{eq:scale-viol-phen}), such that (see
eq.(\ref{eq:scale-viol-new}))

\begin{equation}H_{new}(Q^2)=\int_{0}^{\infty}{dk^2\over k^2} G_{0,new}(k^2/Q^2) A_{{\cal S}}^{new}(k^2)+\Delta
H_{new}(Q^2)\label{eq:delta-H}.\end{equation}
The expansion of $\Delta H_{new}(Q^2)$ is expected to be better convergent than that of $H_{new}(Q^2)$, and can be
dealt with in renormalization scheme invariant way by using the method of effective charges \cite{GG}.

The right hand side of eq.(\ref{eq:scale-viol-phen}) is entirely perturbative. To regularize the renormalons
integrals in the infrared and deal with non-perturbative phenomena (such as power corrections), one can now assume
a non-perturbative modification of the coupling at finite $N_f$, analogous to eq.(\ref{eq:npcouplingQCD}). Its form
should be determined phenomenologically. For instance, in the spirit of the method of effective charges, one could
try a two-point Pad\'{e} interpolation between the (known in perturbation theory)  weak coupling UV form, and the
strong coupling IR form of the non-perturbative  beta function of the effective coupling
$A_{{\cal S},NP}^{new}(k^2)$, assuming a non-perturbative IR fixed point (its value $A_{{\cal S},NP}^{new}(k^2=0)$
is then one of the free parameters to be fitted). $A_{{\cal S},NP}^{new}(k^2)$ could then be determined at {\em
all} scales by integrating its own Gell-Mann-Low like renormalization group equation, and reported into
eq.(\ref{eq:scale-viol-phen}). This method (to be investigated in the near future) does not require the
introduction of any IR cut-off, and represents an interesting alternative (especially if universality of the
Sudakov effective coupling does hold), with a simple physical interpretation, to the shape function approach, as
already remarked in section 2.

\section{Conclusions}
The  IR finite coupling approach has many attractive features in the context of Sudakov
resummation. On the phenomenological  side, we have seen it is able to yield a very simple, yet non-trivial
prediction for the end tail of the Sudakov peak, which goes beyond short distance physics. The parametrization
of the low momentum piece of the Sudakov effective coupling represents also an attractive alternative to the
shape function approach. On the more theoretical side, I have pointed out the rather unusual occurence of a
strongly IR divergent, but nevertheless {\em causal}, effective coupling, in an all order, but still
perturbative, framework. The essential differences in the IR domain between the standard Sudakov anomalous
dimensions (such as cusp), which exhibit completely unphysical behavior at low scales,  and their specific
combinations eq.(\ref{eq:S-coupling}) and (\ref{eq:S-coupling-DY}) called here `Sudakov effective couplings',
where these pathologies cancel out, has been stressed. Only the too strong IR divergence localized at the origin
requires the introduction of a non-perturbative modification. I feel this case is basically different from the
more familiar one of unphysical Landau singularities at {\em finite} distances, and makes  more plausible the
(non-perturbative) IR finite coupling hypothesis. The simplicity of the proposed non-perturbative large $N_f$
ansatzes represents, I believe, further encouragement  to support this speculation. Moreover it was found that the
`natural' resummation of a set of ${\cal O}(N^0)$ terms within the procedure of \cite{Vogelsang} for eikonal cross
sections  remarkably leads in the Drell-Yan case at large $N_f$ to the simplest example (eq.(\ref{eq:A-simple})) of
an IR finite effective Sudakov coupling within perturbation theory itself, with no a-priori need of a
non-perturbative modification.

The IR renormalon and IR finite coupling approaches to power corrections are potentially in conflict with each
other. We have seen it is possible to make consistent these two   approaches, by appropriate use of the
arbitrariness of exponentiated constant terms in Sudakov resummation. This freedom in redefining the Sudakov
exponent bears some connection with alternative forms of Sudakov resummation previously discussed in the
litterature \cite{Beneke-Braun, Catani2}. The latter however involve an infrared cut-off\footnote{The `purely
logarithmic' Sudakov exponent which excludes all constant terms gives an example at large $N_f$ of an alternative
form of Sudakov resummation with a sharp IR cut-off, as we have seen in section 7.}, so that IR renormalons
do not appear through integration over arbitrarily small momenta, but rather through divergences of the
redefined anomalous dimensions (or, as in \cite{Beneke-Braun}, the divergence of some initial condition sitting
outside the exponent). This is in sharp contrast with the present proposal, where integration at arbitrarily small
 momenta is essential, and makes sense
through the notion of IR finite effective coupling. The mathematically simplest solution to resolve the above
mentionned conflict leads to the proposal at large $N_f$ of a `universal' non-perturbative
ansatz for the Sudakov effective coupling (eq.(\ref{eq:npcouplingQCD})). 

Alternatively, it could be that the correct resolution of the  conflict favors one approach over the other, and
that the answer provided by e.g. the `simple' IR finite ansatz eq.(\ref{eq:A-simple}) turns out to be the correct
one (as suggested by its natural occurence in the Drell-Yan process), despite being at odds with the IR renormalon
prediction. These issues should be resolved by a better understanding of OPE at large $N$ in the DIS case. Another
related  question to be settled is the application of the method of \cite{Vogelsang} to DIS, which should determine
the corresponding `natural' Sudakov distribution function and effective coupling. 
Even if the `simple', purely perturbative IR finite effective coupling ansatz turns out to represent an acceptable
answer at the perturbative level at large $N_f$,  there may be extra non-perturbative
effects, similar to those which must occur in the case of the `universal' coupling eq.(\ref{eq:npcouplingQCD}), 
which could also take the form of a non-perturbative modification of the effective coupling at low momenta.
Whatever the correct choice of the Sudakov distribution function turns out to be, 
it remains for 
future phenomenological work  to determine the corresponding  form of the Sudakov effective coupling at
{\em finite} $N_f$ for each process, and test for
eventual deviations from universality.

\acknowledgments
I thank  Yu.L. Dokshitzer, J-P. Lansberg, G. Marchesini, G.P. Salam and G. Sterman for useful discussions. I am
indebted to M. Beneke for reminding me the difficulties of the OPE at large $x$, and for pointing out the
connection of the present findings with alternative forms of Sudakov resummation discussed in the litterature.


\end{document}